\documentclass[aps,superscriptaddress,10pt,nofootinbib,notitlepage,twocolumn,prd]{revtex4-2}

\usepackage{hyperref}
\usepackage{color}
\usepackage{graphicx}
\usepackage{physics}
\usepackage{bm}
\usepackage{amssymb}

\newcommand{\vol}{\mathcal{V}}
\renewcommand{\v}{\vol}

\renewcommand{\b}[1]{\left(#1\right)}

\newcommand{\ba}[1]{\begin{align} #1 \end{align}}
\newcommand{\bes}[1]{\begin{equation}\begin{split} #1 \end{split}\end{equation}}
\newcommand{\bsa}[2]{\begin{subequations}\label{#1}\begin{align} #2 \end{align}\end{subequations}}
\newcommand{\com}{\;,}
\newcommand{\per}{\;.}

\newcommand{\fref}[1]{Fig.~\ref{#1}}
\newcommand{\Fref}[1]{Figure~\ref{#1}}
\newcommand{\tref}[1]{Table~\ref{#1}}

\newcommand{\pref}[1]{(\ref{#1})}
\newcommand{\eref}[1]{Eq.~(\ref{#1})}
\newcommand{\erefs}[2]{Eqs.~(\ref{#1})~and~(\ref{#2})}

\newcommand{\rref}[1]{Ref.~\cite{#1}}
\newcommand{\rrefs}[1]{Refs.~\cite{#1}}

\newcommand{\Mpl}{M_\mathrm{pl}}

\newcommand{\ii}{\mathrm{i}}
\newcommand{\xvec}{{\bm x}}

\newcommand{\kvec}{{\bm k}}

\newcommand{\SM}[1]{\text{\sc sm}}
\newcommand{\KKLT}[1]{\text{$\mathrm{KKLT}$}}
\newcommand{\LVS}[1]{\text{$\mathrm{LVS}$}}
\newcommand{\Vinf}{V_\mathrm{inf}}
\newcommand{\VKKLT}{V_\KKLT{}}
\newcommand{\Hinf}{H_\mathrm{inf}}

\newcommand{\meV}{\ \mathrm{meV}}
\newcommand{\eV}{\ \mathrm{eV}}

\newcommand{\GeV}{\ \mathrm{GeV}}

\begin{document}

\title{
Superheavy Dark Matter from the String Theory Axiverse
}

\author{Siyang Ling}
\affiliation{Department of Physics, City University of Hong Kong,
Tat Chee Avenue, Kowloon, Hong Kong SAR, China}

\author{Andrew~J.~Long}
\affiliation{Department of Physics and Astronomy, Rice University, Houston, TX, 77005, U.S.A.}

\author{Evan McDonough}
\affiliation{Department of Physics, University of Winnipeg, Winnipeg MB, R3B 2E9, Canada}

\author{Alex Hayes}
\affiliation{Department of Physics, University of Winnipeg, Winnipeg MB, R3B 2E9, Canada}
\affiliation{Department of Physics \& Astronomy, University of Manitoba, Winnipeg, MB R3T 2N2, Canada}

\begin{abstract}
We propose heavy axions as a natural superheavy dark matter candidate in string theory, with the relic density of dark matter originating in quantum fluctuations during cosmic inflation. String Theory is well known for the possibility of having tens to hundreds of axion-like particles -- the axiverse. Moduli stabilization generates high-scale masses for many of these, placing them naturally in the `superheavy' regime of particle physics. We consider moduli stabilization in the KKLT framework, featuring a single volume modulus and $C_4$ axion, and a fiducial inflation model minimally coupled to the volume modulus. We demonstrate that both the volume modulus and the axion can be abundantly produced through gravitational particle production. The former is unstable and readily decays to Standard Model particles while the latter (the axion) can be stable and survives to constitute the present day dark matter. 
\end{abstract}

\maketitle

\tableofcontents{}

\section{Introduction}
\label{sec:intro}

A signature prediction of string theory is the existence of a spectrum of axion-like particles, arising from dimensional reduction of extra dimensional gauge fields \cite{Svrcek:2006yi}. In modern parlance this is referred to as the {\it axiverse} of string theory \cite{Arvanitaki:2009fg,Cicoli:2012sz}. While axions were originally proposed in the context of field theory,  as a solution to the strong CP problem \cite{Peccei:1977hh,Wilczek:1977pj,Weinberg:1977ma} and later as a dark matter candidate \cite{Preskill:1982cy,Abbott:1982af,Dine:1982ah}, and while field theory models can also produce a spectrum of fields resembling an axiverse \cite{Maleknejad:2022gyf,Alexander:2023wgk,Alexander:2024nvi}, string theory is distinguished by its prediction for the spectrum of axion masses, decay constants, and couplings to the standard model; see Refs.~\cite{Broeckel:2021dpz,Halverson:2019cmy,Mehta:2020kwu,Mehta:2021pwf,Demirtas:2021gsq,Gendler:2023kjt}.

In parallel, axions and axion-like particles have become a leading candidate for dark matter, with a global experimental effort dedicated to their discovery. Axion dark matter in its original formulation \cite{Preskill:1982cy,Abbott:1982af,Dine:1982ah} is characterized by a coherent scalar field with a sub-eV mass, with the observed relic density originating in an initial misalignment of the field. Following suit, the string theory community has made a considerable effort to understand light and ultralight axions in string theory, including models such as the QCD axion \cite{Demirtas:2021gsq}, fuzzy dark matter \cite{Hu:2000ke,Hui:2016ltb,Cicoli:2021gss,Halverson:2018olu}, and early dark energy \cite{Karwal:2016vyq,Poulin:2018cxd,McDonough:2022pku,Cicoli:2023qri}. The cosmology community has in turn used the string axiverse as motivation for phenomenological studies of ultralight axions.

However, axions in string theory need not be light, and indeed, as we review here, moduli stabilization, namely the process by which the size and shape of extra dimensions are fixed, naturally generates high-scale masses for many of the axions of string theory. Heavy axions were discussed already in the original works on the axiverse, and have been utilized to great success in describing cosmic inflation (see Ref.~\cite{Baumann:2014nda}). In these cases the mass scale of the axion must be far above the ultralight scale of conventional axion dark matter, and is more akin to the realm of {\it superheavy} dark matter.

Superheavy dark matter has undergone its own rise to prominence. Originally popularized as the WIMPZilla \cite{Kolb:1998ki}, superheavy dark matter, characterized by a mass $m \gg $ TeV, exhibits its own rich phenomenology \cite{Carney:2022gse}. Such a dark matter candidate is forbidden by unitarity from having been in thermal equilibrium \cite{Griest:1989wd}, and hence a production mechanism beyond freeze-out is needed to yield the observed relic density. 

A minimal production mechanism for superheavy dark matter is provided by cosmic inflation: the rapid transition in the expansion history of the universe which occurs at the end of inflation can trigger the amplification of quantum vacuum fluctuations into on-shell particles \cite{Parker:1969au, Parker:1971pt}.  Originating with Schr\"odinger \cite{Schrodinger:1939:PVE}, this idea is now referred to as cosmological gravitational particle production (CGPP). See \rrefs{Ford:2021syk,Kolb:2023ydq} for recent reviews of CGPP. The mechanism is essentially the same as that by which the primordial curvature perturbation originates from quantum vacuum fluctuations (e.g. see the discussion in \cite{PhysRevD.15.2738,Lapedes:1977ip,Nakayama:1987dh,Kinney:2003xf}), but also applied to heavy fields at the end of inflation \cite{Kuzmin:1998uv,Chung:1998zb,Chung:1998ua}.  For mass scale near the inflationary Hubble scale, $m\sim H$ and an appropriate reheating temperature, this production can match the observed abundance of dark matter. A growing body of work has shown that CGPP can generate dark matter with a variety of masses and spins \cite{Chung:2004nh,Graham:2015rva,Garny:2015sjg,Ema:2015dka, Ema:2016hlw,Garny:2017kha,Ema:2018ucl,Garny:2018grs,Li:2019ves,Ema:2019yrd,Kolb:2020fwh,Ahmed:2020fhc,Alexander:2020gmv,Redi:2020ffc,Ling:2021zlj,Clery:2021bwz,Dudas:2021njv,Kolb:2021xfn,Kolb:2021nob,Mambrini:2021zpp,Clery:2022wib,Basso:2022tpd,Lebedev:2022vwf,Lebedev:2022cic,Lebedev:2022ljz,Maleknejad:2022gyf,Hashiba:2022bzi,Garcia:2022vwm,Kaneta:2022gug,Kolb:2022eyn,Koutroulis:2023fgp,Kolb:2023dzp,Capanelli:2023uwv,Kaneta:2023uwi,Garcia:2023qab,Garcia:2023awt,Garcia:2023obw,Racco:2024aac,Choi:2024bdn,Capanelli:2024rlk,Capanelli:2024pzd,Jenks:2024fiu,Verner:2024agh,Garcia:2025rut,Feiteira:2025rpe}, 
including masses in the ultralight regime, and in both single field and multifield inflation models. 

With exception of \rref{Allahverdi:2020uax,Allahverdi:2023nov,Leedom:2024qgr}, relatively little attention has been given to the possibility of superheavy dark matter in string theory.  In this work we consider heavy axions in string theory as a superheavy dark matter candidate. We work in the simplest framework for moduli stabilization, featuring a single volume modulus and a single axion. We combine this with an inflation model that is well studied in the CGPP literature, in order to have a basis for comparison and in order to make the first steps in understanding CGPP in string theory. We embed inflation in KKLT following the original proposal in \cite{Kallosh:2004yh} for the minimal coupling between inflation and the volume modulus of KKLT. We numerically compute the CGPP in the combined KKLT+inflation model, along with the associated relic density and conditions for the stability of the dark matter. We note that Ref.~\cite{Leedom:2024qgr}, in a related but distinct analysis, analyzed superheavy axion dark matter produced via preheating. Ref.~\cite{Leedom:2024qgr} differs from the model studied here by a coupling between the inflaton and axion, leading to an oscillating mass term for the latter and particle production.

Despite its relative simplicity, our analysis reveals the structure required of any model in order to realize CGPP of superheavy DM in string theory. Stabilization during inflation bounds the inflationary Hubble scale $H_{\rm inf}$ to be below the scale of the compactification, which naturally implies moduli masses $m_{\rm moduli}$ high above the Hubble scale, $m_{\rm moduli}\gg H$. Conventional CGPP wisdom suggests that the particle production should be exponentially suppressed as $e^{- m/H}$, representing a Boltzmann-like suppression of quantum fluctuations, which, combined with the requirement of stabilization, would preclude CGPP of moduli fields from generating sufficient particles to be dark matter. However this suppression is avoided in inflation models with inflaton mass  $ m_{\rm inf}\gg H$, such as Hilltop models\footnote{While we do not study an explicit string theoretic model of inflation, the hierarchy  $H \lesssim m_{\rm inf}$ is naturally a feature of many string  theory inflation models.}:
Particle production can be efficient in these models for dark matter masses up to and slightly above the inflaton mass scale $m_{\rm inf}$. Thus CGPP of superheavy dark matter in string theory naturally requires a hierarchy $H \ll m_{\rm moduli} \lesssim m_{\rm inf}$.

A second  constraint on model building lies in the stability of the moduli fields. Volume moduli are naturally unstable, owing to their direct coupling, through the K\"{a}hler potential, to all other fields in the theory including the Standard Model. In contrast, superheavy axion fields in string theory can easily be stable: provided that the axion does not couple directly to massless U(1) gauge fields, or do so only weakly (e.g. through loops), the primary decay channel for axions is absent. This is naturally realized in KKLT, where the gauge theory to which the axion directly couples is a confining $SU(N)$ theory. There remains the possibility of decay to gravitons through the Chern-Simons interaction, popularized in the context of Chern-Simons Gravity \cite{Alexander:2009tp} and realized explicitly in closed string axions of Type IIB string theory via their coupling to D-branes \cite{Johnson:2000ch}. We find this imposes a rough upper bound $m_{\rm axion} \lesssim 10^{10}$ GeV, compatible with the estimate in the context of string theory derived in \cite{Leedom:2024qgr}.

Putting these puzzle pieces together, we find that superheavy axions originating in the string theory axiverse are a potentially viable superheavy dark matter candidate, produced via cosmological gravitational particle production. Future work, in the context of an explicit model of inflation in string theory and an explicit realization of the standard model in string theory, will be needed to substantiate this conclusion and we hope that this work motivates future research in this direction.

The structure of this paper is as follows: In Sec.~\ref{sec:axions} we review axions in string theory and in particular heavy axions with masses generated by moduli stabilization, following the recent reviews provided in \cite{Cicoli:2021tzt,Cicoli:2023qri}. In Sec.~\ref{sec:inflation} we build a simplified model of inflation in the KKLT framework for moduli stabilization, elucidate the requirements on the model from successful moduli stabilization, and numerically solve for the cosmological evolution of the fields in the theory. We then proceed to study CGPP, and in Sec.~\ref{sec:production} we numerically solve for the particle production for all three fields in the theory (the inflaton, the volume modulus, and the axion), for  varying model parameters. In Sec.~\ref{sec:darkmatter} we study the axion as a superheavy dark matter candidate, finding a match to the observed relic density and consistency with stability of dark matter. We conclude in Sec.~\ref{sec:conclusion} with a summary of our results and directions for future work.

\section{Heavy Axions In String Theory}
\label{sec:axions}

It is well known that compactifications of string theory's additional spacetime dimensions can lead to as many as tens or hundreds of axion-like particles. Together they comprise the string theory ``axiverse'' \cite{Arvanitaki:2009fg}. Here we focus on Type IIB string theory. In addition to the fundamental axion $C_0$, dimensional reduction of $p$-form gauge fields leads to numerous additional 
axions, defined by \cite{Baumann:2014nda}
\bes{\label{IIBaxions}
    & b^a = \frac{1}{( \alpha')}\int_{\Sigma_a} B_2 
    \com \qquad 
    c^a =  \frac{1}{( \alpha')}\int_{\Sigma_a} C_2 
    \com \\ & \qquad \text{and} \qquad 
    \theta^\alpha = \frac{1}{( \alpha')^2} \int_{D_\alpha} C_4
    \per
}
Here $B_2$ is the Kalb-Ramond two form, $C_2$ and $C_4$ are the Ramond-Ramond two-form and four-form gauge fields, respectively, and $\Sigma_a$ and $D_{\alpha}$ denote a basis of two-cycles and four-cycles of the Calabi-Yau manifold. The number of axions is determined by the topology of the Calabi-Yau manifold, namely the Hodge numbers $h^{1,1}_+$ and $h^{1,1}_-$, which determine the range of $\alpha$ and $a$ respectively.
While some of these axions may be ultralight, allowing for phenomena such as fuzzy dark matter \cite{Hu:2000ke,Hui:2016ltb,Cicoli:2021gss,Halverson:2018olu} and early dark energy \cite{Karwal:2016vyq,Poulin:2018cxd,McDonough:2022pku,Cicoli:2023qri}, many are endowed with superheavy masses as a byproduct of moduli stabilization. 

\subsection{\KKLT{} construction}

The emergence of superheavy axions is most easily seen in the \KKLT{} construction \cite{Kachru:2003aw}. A compactification with $h^{1,1}_+ = 1$ and $h^{1,1}_-=0$ leads to a single 4-form axion $\theta$ and a single volume modulus $\tau$, which together form a K\"ahler modulus $T\equiv \tau + \ii\theta$. The Lagrangian for the scalar sector reads,
\begin{equation}
    {\cal L} = \frac{3 \Mpl^2}{4 \tau^2}(\partial \tau)^2 + \frac{3 \Mpl^2}{4 \tau^2}(\partial \theta)^2 - \VKKLT(\tau,\theta)
    \label{eq:LKKLT}
\end{equation}
where the scalar potential is given by 
\begin{align}\label{eq:VKKLT}
    \VKKLT(\tau,\theta) 
    & = \frac{\mathfrak{a}^2 A^2 e^{-2 \mathfrak{a} \tau/\Mpl}}{6 \Mpl \tau} 
    + \frac{\mathfrak{a} A^2 e^{-2 \mathfrak{a} \tau/\Mpl}}{2 \tau^2} 
    \\ & \quad 
    + \frac{\mathfrak{a} A W_0 e^{-\mathfrak{a} \tau/\Mpl}}{2 \tau^2} \cos(\mathfrak{a} \theta/\Mpl) 
    + \frac{M^2 \Mpl^2}{12 \tau^2} \nonumber 
    \per
\end{align} 
The parameters $\mathfrak{a}$, $A$, $W_0$, and $M$ take real values, and the modulus field is positive $\tau > 0$. 
For $\mathfrak{a} > 0$, $A > 0$, and $W_0 < 0$, the potential's global minimum lies at $(\tau,\theta) = (\tau_0,0)$ for some $\tau_0 > 0$. 
The minimum can be tuned to be (A)dS or Minkowski, with the latter occurring for $W_0$ and $M$ given by 
\begin{align}\label{eq:KKLT_params}
    W_0  \, &= - \frac{1}{3} A \, e^{-\mathfrak{a} \tau_0/\Mpl} \bigl( 5 + 2 \mathfrak{a} \tau_0/\Mpl \bigr)\\
    M \, &= \sqrt{\mathfrak{a}} (A/\Mpl) \, e^{- \mathfrak{a} \tau_0/\Mpl} \sqrt{4 + 2 \mathfrak{a} \tau_0/\Mpl} \nonumber \per
\end{align}
In order to define the mass of the moduli fields, we canonically normalize them as,
\begin{equation}
    \psi = - \sqrt{\frac{3}{2}} \Mpl \log \frac{\tau}{\Mpl} 
    \quad \text{and} \quad 
    \vartheta = \sqrt{\frac{3}{2}} \frac{\Mpl}{\tau} \theta 
    \per \label{eq:norm}
\end{equation}
In terms of these variables there are no mixings in the vacuum and the masses are \footnote{ Note that scaling the KKLT parameters via $\mathfrak{a} \mapsto c^2 \mathfrak{a}$, $A \mapsto A$, $W_0 \mapsto W_0$, $M \mapsto c M$ leads to $\VKKLT \mapsto c^6 \VKKLT$, $\tau_0 \mapsto c^{-2} \tau_0$, $m_\psi \mapsto c^3 m_\psi$ and $m_\vartheta \mapsto c^3 m_\vartheta$. This observation is useful for finding stabilized KKLT parameters giving a desired axion mass. }
\begin{align}\label{eq:mass_formulas}
    m_\psi^2 & = \frac{1}{9\Mpl^4} \mathfrak{a}^3 A^2 e^{-2 \mathfrak{a} \tau_0/\Mpl} (3 + 2 \mathfrak{a} \tau_0/\Mpl) \\ 
    m_\vartheta^2 & = \frac{1}{9\Mpl^4} \mathfrak{a}^3 A^2 e^{-2 \mathfrak{a} \tau_0/\Mpl} (5 + 2 \mathfrak{a} \tau_0/\Mpl) \per \nonumber
\end{align}
Clearly there is no hierarchy of masses. Instead one finds that the mass ratio is 
\begin{equation}
    \frac{m_{\vartheta}^2}{m_{\psi} ^2} = \frac{5 + 2 \mathfrak{a} \tau_0/\Mpl}{3 + 2 \mathfrak{a} \tau_0/\Mpl} 
    \com
\end{equation}
where $\mathfrak{a} \tau_0 > 0$.  
It follows that stabilization of the volume precludes the axion from being ultralight. For the parameters considered in the original KKLT paper, with $A=\Mpl^3$ and $\mathfrak{a}=0.1$, we find $\tau_0\sim 113\Mpl$ and $m_{\psi} \simeq m_\vartheta \simeq 7 \times 10^{-7} \Mpl \simeq 2 \times 10^{12} \; {\rm GeV}$. This mass range, $m\sim 10^{12}$ GeV is more akin to superheavy dark matter \cite{Kuzmin:1998uv,Chung:1998zb,Chung:1998ua,Kolb:1998ki}
than the ultralight regime usually associated with axions. Meanwhile the axion decay constant $f_\vartheta$ can be inferred from the periodicity of the axion potential and is given by 
\begin{equation}
\label{eq:decayconstant}
    f_{\vartheta} = \sqrt{\frac{3}{2}}\frac{\Mpl^2}{\mathfrak{a} \tau_0} \per
\end{equation}
Unlike the axion misalignment mechanism, the decay constant will play no role in setting the relic density. We emphasize that the $f_{\vartheta}$ 
defined above is set by the periodicity of the axion potential, and not by canonical normalization of the axion field, which is defined by Eq.~\eqref{eq:norm}. It follows that the couplings of the axion to other fields, e.g. with gauge fields, are not {\it a priori} given by ${\cal O}(1)f_{\vartheta}^{-1}$, but can vary over a wide range, particularly if the other fields are realized in a distinct sector independent of the physical mechanism used to generate the periodicity of the axion potential used to define the decay constant in the above.

\subsection{Large volume scenario}

Superheavy axions are also expected to arise in the Large Volume Scenario (\LVS{}) for moduli stabilization \cite{Balasubramanian:2005zx,Conlon:2005ki,Cicoli:2008va}. 
In its simplest version the \LVS{} uses two volume moduli, $\tau_s$ and $\tau_b$, which correspond to a small-volume cycle and a big-volume cycle in the compactification. Each modulus is paired with a $C_4$ axion, which we denote by $\theta_s$ and $\theta_b$. The volume of the compactified dimensions is controlled by the big-volume cycle ${\cal V} \sim \tau_b^{3/2}$, where here and in the remainder of this section we set $\Mpl = 1$. The \LVS{} differs from the \KKLT{} construction in that the overall volume is stabilized perturbatively by balancing $\tau_b$ against the energy of $\tau_s$, which is in turn stabilized non-perturbatively as in \KKLT{}. 
The scalar potential is given by \cite{Balasubramanian:2005zx,Conlon:2005ki,Cicoli:2008va} 
\begin{eqnarray}
V_\LVS{}\b{\v,\tau_s,\theta_s} = && \frac{8 \mathfrak{a}_s^2 A_s^2 e^{-2 \mathfrak{a}_s \tau_s} \sqrt{\tau_s}}{3 \v} \\
&&- \frac{4 \mathfrak{a}_s A_s \tau_s |W_0|\,e^{-\mathfrak{a}_s \tau_s}}{\v^2}\cos\b{\mathfrak{a}_s \theta_s}  \nonumber \\
&& + \frac{3|W_0|^2\hat\xi}{4\v^3} +\frac{M^2}{\v^{4/3}} \, \nonumber ,
\label{VLVS1}
\end{eqnarray}
where $\mathfrak{a}_s$, $A_s$ describe a stack of branes on the cycle parametrized by $\tau_s$ and $\hat{\xi}$ is a perturbative correction generated by higher-derivative curvature terms \cite{Becker:2002nn}.

At this level the potential is independent of $\theta_b$ and hence the $\theta_b$ axion remains massless. A small mass for $\theta_b$ can be generated non-perturbatively, but is not needed for the stabilization of $\tau_b$. The $\theta_s$ axion, on the other hand, is analogous to $\theta$ of \KKLT{}, and receives a large mass from the stabilization of $\tau_s$. 
This straightforwardly generalizes to \LVS{} with an arbitrary number of moduli,  $\{ \tau_{s_i},\theta_{s_i} \}$ with $i=1...(h^{1,1}_+-1)$, where now the scalar potential is simply a sum over $i$. All $C_4$ axions are heavy, except for the single remaining $\theta_b$ axion.

\subsection{Two-Form Axions}

The string theory axiverse includes several additional axions besides the ones associated with the \KKLT{} and \LVS{} moduli stabilization schemes. 
Among these are the $b$ and $c$ axions resulting from dimensional reduction of the two-forms $B_2$ and $C_2$. Their masses arise in qualitatively different ways: the shift-symmetry of the $B_2$ axion is broken already in perturbative string theory, whereas the shift-symmetry of the $C_2$ axion is broken only by effects such as instantons.

These two-form axions are incorporated into the 4d theory via the K\"ahler modulus $T$ as
\begin{equation}
   T = \tau + \ii \theta -\frac14 g_s k G(G+\bar{G})
\end{equation}
where $g_s$ is the string coupling and $k$ is a triple-intersection number of the Calabi-Yau manifold.
The axions, $b$ and $c$, reside in $G$ as
\begin{equation}
    G  = \frac{b}{g_s}+\ii(c-C_0 b) 
    \com
\end{equation}
and for simplicity we have fixed $h^{1,1}_-=1$. The volume takes the standard form
\begin{equation}
\mathcal{V} = \frac13\sqrt{\frac{2}{\tilde{k}}}\,\tau^{3/2} \com
\end{equation}
where $\tilde{k}$ is again a triple-intersection number, but now the volume modulus is shifted as,
\begin{equation}
\tau = \Re(T) + \frac{k}{2g_s}\,b^2 \com
\label{Vodd}
\end{equation}
which explicitly depends on $b$. This relation inextricably links $b$ to the stabilization of the volume, and leads to a large mass for the $B_2$ axion $b$. In addition, a mass for $b$ can be generated by D-terms \cite{Cicoli:2023qri}. 
The $C_2$ axion $c$ remains as a light degree of freedom, and is suitable candidate for fuzzy dark matter and Early Dark Energy, with a mass that can be generated by a variety of non-perturbative mechanisms.

\subsection{Fundamental axion}

Returning to the fundamental axion $C_0$, we note that high-scale stabilization of the axio-dilaton is implicit in the starting point of both \LVS{} and \KKLT{}, where the complex structure moduli are stabilized by the the flux-induced superpotential $W_0$, following earlier work \cite{Giddings:2001yu}. Explicitly,
\begin{equation}
W_0 = \int \left( F_3 - S H_3 \right) \wedge \Omega\,,
\end{equation}
where $S \equiv e^{-\phi}+{\rm i}C_0$ is the axio-dilaton, $F_{3}$ and $H_3$ are the field strengths of $C_2$ and $B_2$ respectively, and $\Omega$ is the holomorphic 3-form of the Calabi-Yau manifold. This superpotential generates a mass for the dilaton $\phi$ and in so doing generates a comparable mass for the fundamental axion $C_0$.  A heavy dilaton, motivated, e.g., to satisfy constraints from fifth force experiments, therefore implies a heavy $C_0$ axion. We note that models where both the axion and dilaton are light and are consistent with data have been proposed in e.g.~\rref{Burgess:2021qti,Burgess:2021obw,Brax:2022vlf,Brax:2023qyp}.

\section{Inflation in KKLT}
\label{sec:inflation}

We now turn to cosmology. For simplicity, we adopt the KKLT scenario for moduli stabilization. Instead of building an explicit model of inflation in string theory, we adopt the approach considered by two of the authors of KKLT in Ref.~\cite{Kallosh:2004yh}, wherein the inflaton is coupled to the volume modulus only by a power of the overall volume. This constitutes the minimally coupled inflation model.

\subsection{Model of inflation}

Concretely, we consider a model defined by Lagrangian \cite{Kallosh:2004yh}\footnote{Here the power of $\tau^{-2}$ vs $\tau^{-3}$ is due to warping \cite{Kachru:2003sx}.}
\begin{equation}
    {\cal L}  = {\cal L}_\mathrm{kin} - \VKKLT{}(\tau,\theta) - \left( \frac{\tau_0}{\tau} \right)^{2} \, \Vinf(\phi)
    \label{eq:Lfull}
\end{equation}
where $\Vinf(\phi)$ defines the inflaton potential normalized at $\tau=\tau_0$, and $\VKKLT$ is the KKLT potential given by \eref{eq:VKKLT}. The kinetic terms ${\cal L}_\mathrm{kin}$ are given by the KKLT kinetic terms of \eref{eq:LKKLT} and a canonical kinetic term for $\phi$. 
Note that the additional term does not introduce a mixing between $\phi$ and $\tau$ in the vacuum, and the mass relations in \eref{eq:mass_formulas} are unchanged. 

Heuristically, the minimal coupling of the inflation to KKLT emerges from promoting the KKLT uplift parameter $M$ to a function of the inflaton. This can be explicitly and straightforwardly realized in supergravity constructions of KKLT, see e.g. \cite{Ferrara:2014kva,McDonough:2016der,Kallosh:2017wnt}. A minimal example, which reproduces \eref{eq:Lfull}, is given in App. \ref{app:sugra}.

\begin{figure}
    \centering
    \includegraphics[width=0.9\linewidth]{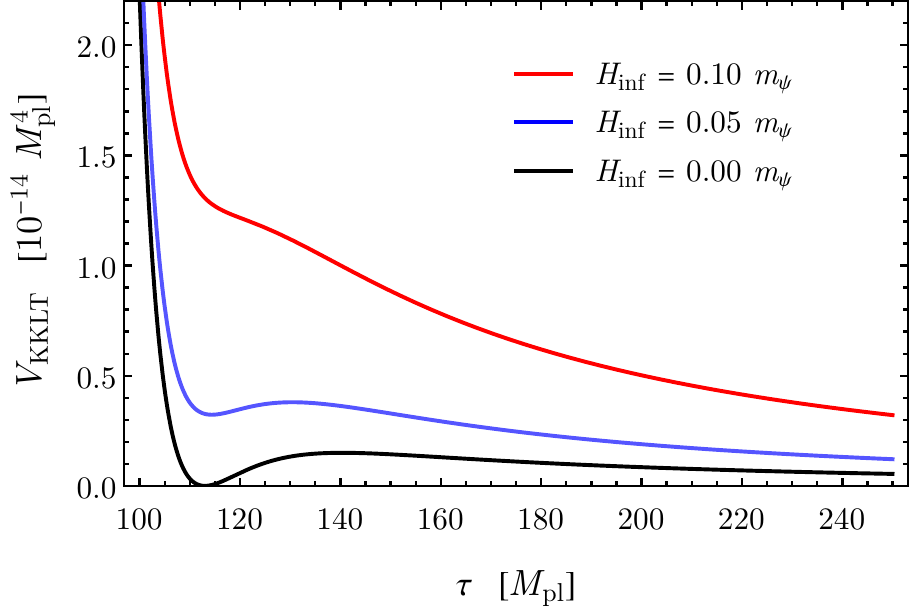}
    \caption{Destabilization of the \KKLT{} volume modulus $\tau$ due to inflation. We show the \KKLT{} potential minimally coupled to inflation as per \eref{eq:Lfull}, for several values of $\Hinf/m_{\psi}$ where $\Hinf$ is the inflationary Hubble scale and $m_{\psi}$ is the mass of canonically normalized volume modulus. Stabilization of the volume modulus requires $\Hinf \ll m_{\psi}$. Parameters are $\mathfrak{a} = 0.1$, $A = \Mpl^3$, $\tau_0 = 113 \Mpl$, and $W_0$, $M$, determined by \eref{eq:KKLT_params}.  
    }
    \label{fig:destab}
  \end{figure}

As emphasized in Ref.~\cite{Kallosh:2004yh}, the minimal coupling \pref{eq:Lfull} presents a serious risk of destabilization. Heuristically, if the inflationary vacuum energy $\Vinf \approx 3 \Mpl^2 \Hinf^2$ is larger than the barrier to decompactification, then the volume modulus $\tau \sim \psi$ is no longer stabilized and will roll to infinity. This in turn bounds the energy scale of inflation to be well below the energy scale of the compactification, and by extension, the mass of the volume modulus. This implies a hierarchy between the inflationary Hubble scale and the mass of the (canonically normalized) volume modulus: 
\begin{equation}
    \Hinf \ll m_{\psi} 
    \per
\end{equation}
We emphasize this point by plotting in \fref{fig:destab} the potential as a function of the volume modulus for varying $\Hinf$ and $m_{\psi}$, analogous to Figure 2 of Ref.~\cite{Kallosh:2004yh}. One may appreciate that, for this choice of parameters, the volume modulus is destabilized for $\Hinf \gtrsim 0.1 m_{\psi}$.  
Since the volume modulus $\tau \sim \psi$ and axion $\theta \sim \vartheta$ have similar masses, avoiding destabilization requires $m_{\psi} \approx m_{\vartheta} \gtrsim 10 \Hinf$.

On the other hand, in this work we are expressly interested in inflationary production of string theory moduli, such as the volume modulus and its associated axion. The conventional wisdom is that cosmological gravitational particle production quickly becomes inefficient for particle masses above the scale of inflation.  
The stabilization condition $m_{\psi} \approx m_{\vartheta} \gtrsim 10 \Hinf$ ostensibly precludes any non-negligible production of string theory moduli.
However, for models in which there is a hierarchy between the inflationary Hubble scale $\Hinf$ and the inflaton mass $m_\phi$, gravitational particle production can still be efficient if the moduli masses are in the range 
\begin{equation}
    \Hinf \ll (m_\psi \approx m_{\vartheta}) \lesssim m_{\phi}
    \;,
\end{equation}
or even if they are just a factor of $2$ or $3$ larger than $m_\phi$.  
This regime has been studied in the context of hilltop inflation \cite{Ema:2018ucl,Basso:2022tpd,Kolb:2023dzp}. 
In the string theory context, many models have this feature.

To simplify the analysis and connect with the existing particle production literature, we focus on a fiducial Hilltop inflation model with potential given by 
\begin{equation}
    \Vinf(\phi) = \frac{m_\phi^2 v^2}{72} \biggl( 1 - \frac{\phi^6}{v^6} \biggr)^{\!\!2}
    \label{eq:Vinf}
\end{equation}
We identify the inflationary Hubble scale as $\Hinf = \sqrt{\Vinf(0) / 3 \Mpl^2} = m_\phi v / \sqrt{216} \Mpl$.  
This model is compatible with Planck 2018 CMB constraints and features a hierarchy $m_{\phi} / \Hinf \approx 29.4$ for $v = 0.5 \Mpl$.

\begin{table}[t]
\caption{Model parameters and initial conditions for background evolution.  We normalize dimensionful quantities to the reduced Planck mass $\Mpl \approx 2.435 \times 10^{18} \GeV$.  The parameters used for numerics have higher precision, $W_0$ and $M$ are calculated using \eref{eq:KKLT_params}, and $H_e$ has an order $1\%$ level dependence on the KKLT parameters (not shown).  }
\label{tab:parameters}
\begin{center}
  Hilltop inflation parameters:
  \begin{tabular}{|c||c|c|c|c|}\hline
    & $m_\phi~[\Mpl]$ & $v~[\Mpl]$ & $\Hinf~[\Mpl]$ & $H_e~[\Mpl]$ \\\hline
        P1,P2 & $1.7\times 10^{-6}$ & $0.5$ & $5.78\times 10^{-8}$ & $5.42 \times 10^{-8}$ \\ \hline
    P3 & $3.0\times 10^{-9}$ & $0.2$ & $4.08\times 10^{-11}$ & $3.97\times 10^{-11}$ \\ \hline
  \end{tabular} \\ \phantom{.} \\ 
  KKLT potential parameters:
  \begin{tabular}{|c||c|c|c|c|}\hline
    & $\mathfrak{a}$ & $A~[\Mpl^3]$ & $W_0~[\Mpl^3]$ & $M~[\Mpl^2]$  \\\hline
    P1 & $0.109$ & $2.14$ & $-0.000901$ & $0.000175$ \\ \hline
    P2 & $0.0432$ & $2.14$ & $-0.000901$ & $0.00011$ \\ \hline
    P3 & $0.0628$ & $0.02$ & $-7.01\times 10^{-7}$ & $9.44\times 10^{-8}$ \\ \hline
  \end{tabular} \\ \phantom{.} \\ 
  \begin{tabular}{|c||c|c|c|}\hline
    &  $\tau_0~[\Mpl]$ & $m_\psi~[\Mpl]$ & $m_\vartheta~[\Mpl]$ \\\hline
    P1 & $90.9$ & $6.23\times 10^{-6}$ & $6.49\times 10^{-6}$ \\ \hline
    P2 & $229.$ & $1.56\times 10^{-6}$ & $1.62\times 10^{-6}$ \\ \hline
    P3 & $200.$ & $1.94\times 10^{-9}$ & $2.01\times 10^{-9}$ \\ \hline
  \end{tabular} \\ \phantom{.} \\ 
  Initial conditions for the background fields:
  \begin{tabular}{|c||c|c|c|c|}
    \hline
    & $\bar{\phi}_i~[\Mpl]$ & $\dot{\bar{\phi}}_i~[\Mpl^2]$ & $\bar{\tau}_i~[\Mpl]$ & $\dot{\bar{\tau}}_i~[\Mpl^2]$ \\ \hline
    P1 & $0.0483$ & $1.16\times 10^{-11}$ & $90.9$ & $-7.39\times 10^{-17}$ \\ \hline
    P2 & $0.0483$ & $1.16\times 10^{-11}$ & $229.$ & $-1.86\times 10^{-16}$ \\ \hline
    P3 & $0.00376$ & $5.82\times 10^{-18}$ & $200.$ & $-3.12\times 10^{-25}$ \\ \hline
  \end{tabular}
\end{center}
\end{table}

\subsection{Background Evolution}

To study the dynamics of the three scalar fields during inflation and soon after inflation, we need their equations of motion as well as the equation for the spacetime metric.    
We assume a homogeneous, isotropic, and spatially-flat FLRW spacetime, and we write the spacetime interval as $(\dd s)^2 = (\dd t)^2 - a(t)^2 |\dd \xvec|^2$ where $t$ is the time coordinate, $\xvec$ is the comoving spatial coordinate 3-vector, $a(t)$ is the scale factor, and $H(t) = (1/a) (da/dt)$ is the Hubble parameter. 
In this section we restrict our attention to homogeneous field configurations, denoted by $\bar{\phi}(t)$, $\bar{\tau}(t)$, and $\bar{\theta}(t)$; we consider inhomogeneities associated with quantum fluctuations in the next section. 
The Lagrangian \pref{eq:Lfull} gives rise to the equations of motion for the three scalar fields.  
The equation for the axion is solved by $\bar{\theta}(t) = 0$.  
With this restriction the equations for the inflaton and volume modulus take the form 
\begin{subequations}\label{eq:background_eom}
\ba{
    0 & = 
    \ddot{\bar{\phi}} 
    + 3 H \dot{\bar{\phi}} 
    - \frac{m^2 v}{6} \frac{\tau_0^2}{\bar{\tau}^2} \biggl( 1 - \frac{\bar{\phi}^6}{v^6} \biggr) \frac{\bar{\phi}^5}{v^5} \\ 
    0 & = 
    \ddot{\bar{\tau}} 
    + 3 H \dot{\bar{\tau}} - \frac{(\dot{\bar{\tau}})^2}{\bar{\tau}} 
    - \frac{2 \mathfrak{a}^2 A^2}{9 \Mpl^4} \mathfrak{a} \bar{\tau} e^{-2 \mathfrak{a} \bar{\tau}/\Mpl} 
    \\ & 
    - \frac{7 \mathfrak{a}^2 A^2}{9 \Mpl^3} e^{-2 \mathfrak{a} \bar{\tau}/\Mpl} 
    - \frac{2 \mathfrak{a}^2 A^2}{3 \Mpl^2} \frac{e^{-2 \mathfrak{a} \bar{\tau}/\Mpl}}{\mathfrak{a} \bar{\tau}} 
    - \frac{\mathfrak{a}^2 A W_0}{3 \Mpl^3} e^{-\mathfrak{a} \bar{\tau}/\Mpl} \nonumber \\ & 
    - \frac{2 \mathfrak{a}^2 A W_0}{3 \Mpl^2} \frac{e^{-\mathfrak{a} \bar{\tau}/\Mpl}}{\mathfrak{a} \bar{\tau}}
    - \frac{\mathfrak{a} M^2}{9} \frac{1}{\mathfrak{a} \bar{\tau}} 
    - \frac{m^2 v^2 \tau_0}{54 \Mpl^2} \frac{\tau_0}{\bar{\tau}} \biggl( 1 - \frac{\bar{\phi}^6}{v^6} \biggr)^{\!\!2} \nonumber
    \com
}
where dots denote $d/dt$.  
Notice how interactions between these two fields couple their equations of motion. 
The equation of motion for the metric is the first Friedmann equation, which takes the form 
\ba{
    3 \Mpl^2 H^2 & = \frac{1}{2} (\dot{\bar{\phi}})^2 + \frac{3}{4} \Mpl^2 \frac{(\dot{\bar{\tau}})^2}{\bar{\tau}^2} \\
    & + \VKKLT{}(\bar{\tau},0) + \frac{\tau_0^2}{\bar{\tau}^2} \, \Vinf(\bar{\phi}) \nonumber
    \com
}
\end{subequations}
where $H = \dot{a}/a$ and where $\VKKLT{}$ and $\Vinf$ are given by \erefs{eq:VKKLT}{eq:Vinf} respectively. 
Collectively \eref{eq:background_eom} governs the evolution of the homogeneous inflaton field $\bar{\phi}(t)$, modulus field $\bar{\tau}(t)$, and scale factor $a(t)$ during inflation and after inflation. 
This approach neglects decays of the inflaton and modulus fields, which would cause their field amplitudes to approach zero exponentially quickly when the age of the universe reaches their lifetimes, and would cause the universe to become radiation dominated.  
Provided that particle production is completed before these decays occur, called ``late-reheating regime'' in \rref{Kolb:2023dzp}, our approach is expected to be a good approximation.

To study the dynamics of the inflaton and modulus fields during inflation we numerically solve the background equations of motion \eref{eq:background_eom} for the model parameters given in \tref{tab:parameters}.  
Parameter sets P1 and P2 are chosen to yield inflationary observables ($A_s$ and $n_s$) that are consistent with observations, while parameter set P3 is chosen to yield a cosmologically-long lived axion. 
Slow roll inflation is possible by selecting an initial condition with $0 < \bar{\phi}_i \ll v$ such that the inflaton field $\phi$ is near the hilltop at $\phi = 0$.  
For $\phi = 0$ the potential in \eref{eq:Lfull} is minimized at 
\bes{
    \tau \approx \tau_0 + \frac{m^2 v^2 \Mpl^2 e^{2 \mathfrak{a} \tau_0 / \Mpl}}{2 \mathfrak{a}^3 A^2 (3 + 2 \mathfrak{a} \tau_0 / \Mpl) + m^2 v^2 \Mpl^2 e^{2 \mathfrak{a} \tau_0 / \Mpl}} \frac{\tau_0}{3} 
    ,
}
where we have used \eref{eq:KKLT_params}. 
If we initialize $\tau$ nearby to this value then it remains approximately constant throughout most of inflation until $\phi$ grows comparable to $v$.  
It is reasonable to set $\tau$ to its minimum, since this field is heavier than Hubble ($m_\tau > \Hinf$) and it evolves into its minimum quickly near the beginning of inflation.

For the model specified by parameter set P1 in \tref{tab:parameters}, our results appear in \fref{fig:background}. 
We show the trajectory of the inflaton field $\bar{\phi}(t)$ and the modulus field $\bar{\tau}(t)$ as a function of the FLRW scale factor $a(t)$ during the last few $e$-foldings of inflation and after inflation. 
During inflation, the modulus field remains approximately constant and the inflaton field grows as it slowly rolls down the hilltop potential. Inflation ends when the fields approach the local minimum at $(\tau,\phi) = (\tau_0,v)$. 
After inflation the fields oscillate about the minimum of the potential, and the field amplitudes decrease as $a^{-3/2}$ as a consequence of the cosmological expansion. 

\subsection{Approximately single-field dynamics and perturbations}

Although this model of inflation involves three scalar fields --- the inflaton $\phi$, the modulus $\tau$, and the axion $\theta$ --- only two of them are dynamical during inflation, since the axion is a spectator $\bar{\theta}(t) = 0$.  
While the inflaton rolls from $\phi \approx 0$ to $\phi \approx v$, the modulus field $\bar{\tau}(t)$ varies by only $\approx 0.04\%$ over the course of inflation. 
Note that this slight shift in $\bar{\tau}$ is an example of moduli vacuum misalignment \cite{Cicoli:2016olq}.  
 This observation indicates that the background dynamics are effectively single-field, namely that of the Hilltop inflation model.  

To further quantify the effectively single-field nature of the background evolution, we calculate the pseudoscalar turn rate $\omega$.  
Following the standard treatment of multifield inflation \cite{Gordon:2000hv,Wands:2002bn,Langlois:2008mn,Peterson:2010np,Gong:2011uw,Kaiser:2012ak,Gong:2016qmq}, we describe the instantaneous direction in field space along which the system evolves using the unit vector
\begin{equation}
   \hat{\sigma}^I(t) \equiv \dot{\Phi}^I / |\dot{\Phi}^I| 
   \com
\end{equation}
where $\Phi^1 = \bar{\phi}(t)$ and $\Phi^2 = \bar{\tau}(t)$.  
The rate of change of $\hat{\sigma}^I$ determines a vector turn rate,  $ \omega^I \equiv {\cal D}_t \hat{\sigma}^I$, where ${\cal D}_t$ denotes a directional covariant derivative on field space, and from this one determines  the pseudoscalar turn rate 
$\omega \equiv \epsilon_{IJ} \, \hat{\sigma}^I \, \omega^J$ \cite{McDonough:2020gmn}. 
For parameter set P1 in \tref{tab:parameters}, we calculate the turn rate $\omega(t)$ using our numerical solution for the background field evolution, and we present the result in \fref{fig:isocurvature_mass}. 
We find that the turn rate is much smaller than the Hubble expansion rate at all times during inflation, indicating effectively single field dynamics.

\begin{figure}[t]
\centering
\includegraphics[width=0.50\textwidth]{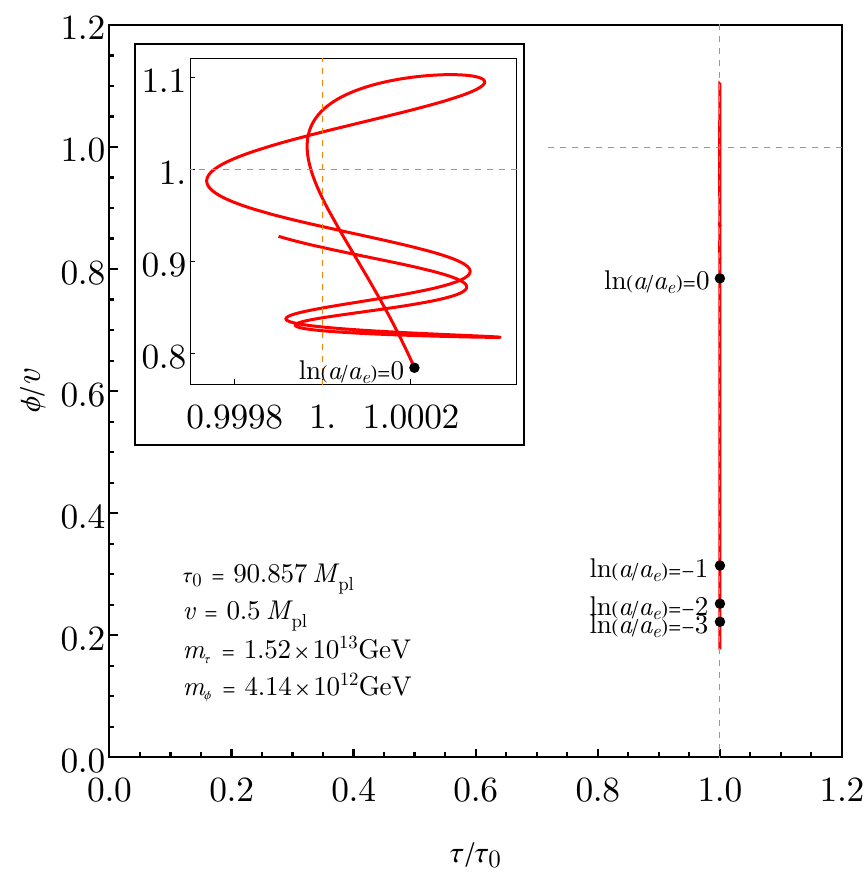}
\caption{\label{fig:background}
Evolution of the inflaton and modulus field backgrounds near the end of inflation and after inflation. 
The red curve shows the solution trajectory, and the black dots mark several $e$-foldings before and after the end of inflation, which occurs at $a=a_e$. 
This numerical calculation is performed using parameter set P1 in \tref{tab:parameters}, and the solution is qualitatively similar for other parameter choices.   
}
\end{figure}

\begin{figure}
\centering
\includegraphics[width=0.50\textwidth]{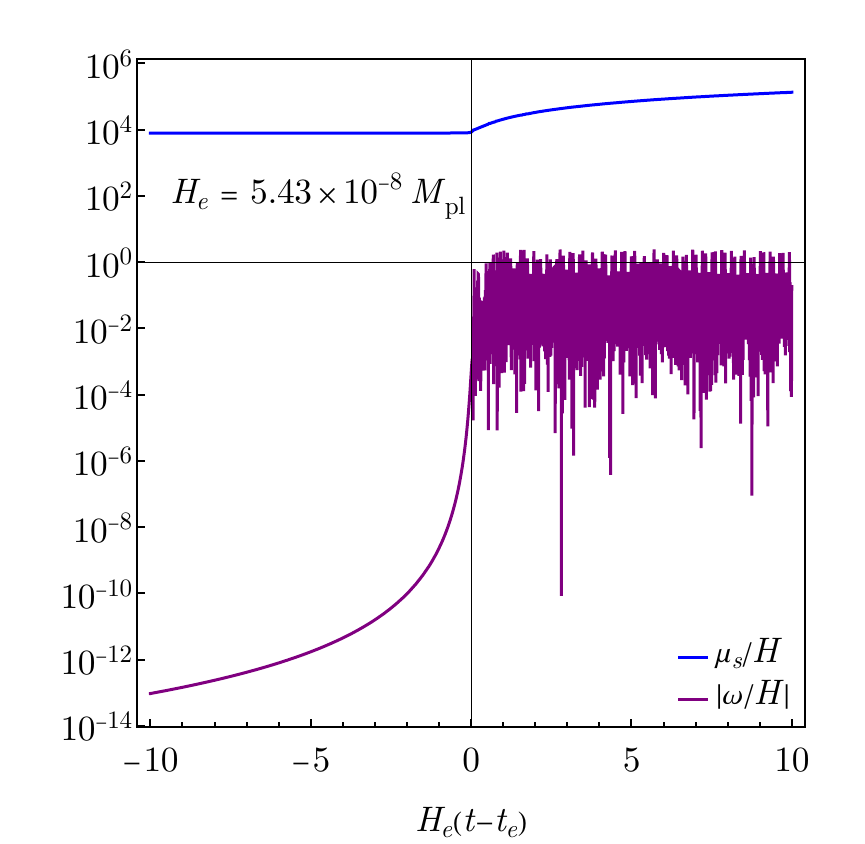}
\caption{\label{fig:isocurvature_mass}
   Isocurvature mass $\mu_s(t)$ and turn rate $\omega(t)$ compared to $H(t)$ during inflation ($t < t_e$) and after inflation ($t > t_e$). 
   The model parameters are given by P1 in \tref{tab:parameters}.
}
\end{figure}

To further justify the single-field nature of this model, we calculate the isocurvature mass.  
Inhomogeneities in $\phi$ and $\tau$ can be expressed using a basis of curvature perturbations ${\cal R}_\kvec(t)$ and isocuvature perturbations ${\cal S}_\kvec(t)$. 
These correspond to the fluctuations in the (instantaneous) parallel and orthogonal directions to the field space trajectory.  
A review of cosmological perturbations in multifield inflation can be found in e.g.~ Refs.~\cite{Gordon:2000hv,Wands:2002bn,Langlois:2008mn,Peterson:2010np,Gong:2011uw,Kaiser:2012ak,Gong:2016qmq}. 
On large scales the equations of motion for the curvature and isocurvature perturbations take on a simple form,
\begin{align}
\label{eq:dotRS}
    & \dot{\cal R}_\kvec = 2 \omega {\cal S}_\kvec \\ 
    & \ddot{\cal S}_\kvec + (3 + \delta) H \dot{\cal S}_\kvec +  \mu^2 _s {\cal S}_\kvec = 0 
    \com
\end{align}
where $\mu_s^2$ is the mass of superhorizon isocurvature perturbations, given by
\begin{equation}
\label{eq:mus}
    \mu_s^2(t) = {\cal M}_{ss} + 3 \omega^2 + H^2(2 \epsilon-\eta)(3+5 \epsilon-\eta)+H^2 \eta \kappa ,
\end{equation}
and the slow-roll parameters are denoted by $\epsilon$, $\eta$, $\delta$, and $\kappa$.  
Here ${\cal M}_{ss}$ is the isocurvature component of the mass matrix ${\cal M}^I_{\;J}$, with the former defined as ${\cal M}_{ss} \equiv ({\cal G}^{IJ} - \hat{\sigma}^I \hat{\sigma}^J ) {\cal M}_{IJ}$ and the latter as
${\cal M}^I_{\;J} \equiv {\cal G}^{IK} {\cal D }_J {\cal D}_K V - {\cal R}^{I} \,_{LMJ} \dot{\Phi}^L \dot{\Phi}^M$.

In the present model, with inflation proceeding primarily along the $\phi$ direction  (see \fref{fig:background}), the isocurvature direction can be identified with $\tau$, and the curvature direction with $\phi$. The lack of any turning in the field space trajectory implies, by \eref{eq:dotRS}, that the curvature and isocurvature perturbations are decoupled on large scales. Moreover,  since $m_{\psi} \gg \Hinf$ (where $\psi$ is the canonically normalized $\tau$ field), in order to avoid destabilization of the volume modulus, one may infer that the isocurvature perturbation is heavy, and will decay outside the horizon. Together these features imply a negligible modification to the predictions of the single-field hilltop inflation model. 

To substantiate this 
we calculate the isocurvature mass $\mu_s(t)$ and plot the result in Fig.~\ref{fig:isocurvature_mass} alongside the turn rate $\omega(t)$.  
One may appreciate that $\mu_s^2 \gg  H^2$ at all times, indicating that the isocurvature perturbations are indeed heavy and hence will decay on superhorizon scales. Meanwhile the turn rate is negligible at all times during inflation. Taken in conjunction, this implies that cosmological perturbations on CMB scales are effectively single-field.
Combined with the single field background evolution, one may infer that the KKLT+inflation model \eref{eq:Lfull} inherits the CMB predictions of the underlying single-field Hilltop inflation model, which are in agreement with Plank 2018 measurements of the inflationary observables for the parameter sets in \tref{tab:parameters}.

\section{Particle Production}
\label{sec:production}

Now we turn our attention to cosmological gravitational particle production 
of the modulus and axion fields.  
For quantum fields that are not conformally coupled to gravity, the cosmological expansion in a homogeneous and isotropic FLRW spacetime leads to particle production~\cite{Parker:1968mv,Parker:1969au,Parker:1971pt}.  
See the review articles \cite{Ford:2021syk,Kolb:2023ydq} for additional details and original references.
We use the Bogolubov formalism to calculate the spectrum and abundance of these particles that are created during inflation and at the end of inflation.  See also \rref{Leedom:2024qgr} for a study of string axion production through preheating at the end of inflation.

\subsection{Mode Equations and Particle Number Density}

To study the dynamics of the fields' inhomogeneities we employ a system of coupled mode equations.  
Let the inflaton field $\phi(t,\xvec)$, modulus field $\tau(t,\xvec)$, and axion field $\theta(t,\xvec)$ be decomposed as 
\bsa{eq:perturbation_eqns}{
    \phi(t,\xvec) & = \bar{\phi}(t) + \int \! \! \frac{\dd^3 \kvec}{(2\pi)^3} \, \phi_\kvec(t) \, e^{\ii \kvec \cdot \xvec} \\ 
    \tau(t,\xvec) & = \bar{\tau}(t) + \int \! \! \frac{\dd^3 \kvec}{(2\pi)^3} \, \tau_\kvec(t) \, e^{\ii \kvec \cdot \xvec} \\ 
    \theta(t,\xvec) & = \bar{\theta}(t) + \int \! \! \frac{\dd^3 \kvec}{(2\pi)^3} \, \theta_\kvec(t) \, e^{\ii \kvec \cdot \xvec} 
    \com
  }
where $\bar{\phi}(t)$, $\bar{\tau}(t)$, and $\bar{\theta}(t)$ are the homogeneous inflationary background solutions, and where $\phi_\kvec(t)$, $\tau_\kvec(t)$, and $\theta_\kvec(t)$ are the amplitudes of the Fourier mode with comoving wavevector $\kvec$.  
Recall that $\bar{\theta}(t) = 0$ whereas $\bar{\phi}(t)$ and $\bar{\tau}(t)$ satisfy \eref{eq:background_eom}. 
We derive the equations of motion for the Fourier mode amplitudes, which are found to be 
\bsa{eq:mode_equations}{
    0 & = 
    \ddot{\phi}_\kvec 
    + 3 H \dot{\phi}_\kvec 
    + \left( \frac{k^2}{a^2} + \frac{\tau_0^2}{\bar{\tau}^2} \, \Vinf''(\bar{\phi}) \right) \phi_\kvec \\
    & \quad - \frac{2 \tau_0^2}{\bar{\tau}^3} \, \Vinf'(\bar{\phi}) \, \tau_\kvec  \nonumber
    \\ 
    0 & = 
    \ddot{\tau}_\kvec 
    + \left( 3 H - \frac{2 \dot{\bar{\tau}}}{\bar{\tau}} \right) \dot{\tau}_\kvec
    - \frac{4 \tau_0^2}{3 \Mpl^2 \bar{\tau}} \, \Vinf'(\bar{\phi}) \, \phi_\kvec 
    \\ & \quad 
    +  \Biggl(
        \frac{k^2}{a^2}
        - 6 H \frac{\dot{\bar{\tau}}}{\bar{\tau}}
        + \frac{3 (\dot{\bar{\tau}})^2}{\bar{\tau}^2}
        - 2 \frac{\ddot{\bar{\tau}}}{\bar{\tau}} 
        + \frac{4\tau_0^2}{\Mpl^2\bar{\tau}^2} \Vinf(\bar{\phi}) 
        \nonumber \\ & \quad \qquad 
        + \frac{2 \bar{\tau}^2}{3\Mpl^2} \partial_\tau^2 \VKKLT{}(\bar{\tau}, \bar{\theta})
        \Biggl) \tau_\kvec 
    \nonumber \\
    0 & = 
    \ddot{\theta}_\kvec
    + \left( 3 H - \frac{2 \dot{\bar{\tau}}}{\bar{\tau}} \right) \dot{\theta}_\kvec \\
    & \quad + \left(\frac{k^2}{a^2} + \frac{2 \bar{\tau}^2}{3\Mpl^2} \partial_\theta^2 \VKKLT{}(\bar{\tau}, \bar{\theta}) 
    \right) \theta_\kvec  \nonumber
    \com
}
where $k = |\kvec|$. 
In this derivation we have retained only the linear terms, and consequently each Fourier mode (labeled by $\kvec$) evolves independently of the others.  
Note also that at the linear order $\phi_\kvec$ and $\tau_\kvec$ are coupled while $\theta_\kvec$ evolves independently. 
We set to zero the scalar metric perturbations, which would only appear in the equations for $\phi_\kvec$ and $\tau_\kvec$ at linear order, but which are negligible for modes that remain inside the horizon ($k > a_e H_e$). %

We solve the mode equations \pref{eq:mode_equations} as an initial value problem using numerical methods. 
We set an initial condition at time $t_i$, which is a few $e$-foldings before the end of inflation. 
For each of the three Fourier mode amplitudes, we change variables using 
\bsa{}{
    \phi_\kvec(t) & = a(t)^{-3/2} \, \chi_\kvec^{(\phi)}(t) \\ 
    \tau_\kvec(t) & = \sqrt{\frac{2}{3}} \, a(t)^{-3/2} \, \frac{\bar{\tau}(t)}{\Mpl} \, \chi_\kvec^{(\tau)}(t) \\ 
    \theta_\kvec(t) & = \sqrt{\frac{2}{3}} \, a(t)^{-3/2} \, \frac{\bar{\tau}(t)}{\Mpl} \, \chi_\kvec^{(\theta)}(t) 
    \com
}
such that the $\chi_\kvec^{(\cdot)}(t)$ have canonically-normalized kinetic terms.  
For each mode function we impose the Bunch-Davies initial condition at time $t_i$: 
\bes{
  \chi_\kvec^{(\cdot)}(t_i) & = \sqrt{\frac{\pi}{4 H(t_i)}} H_0\left(\frac{k}{a(t_i) H(t_i)}\right) \\
  \dot{\chi}_\kvec^{(\cdot)}(t_i) & = \sqrt{\frac{\pi}{4 H(t_i)}} H_1\left(\frac{k}{a(t_i) H(t_i)}\right) \frac{k}{a(t_i)} 
  \;,
}
where $H_n(z)$ is the $n^\mathrm{th}$ Hankel function of the first kind. 
Having solved for the evolution, we evaluate the Bogolubov coefficients for each field using \eref{eq:betak_sq} and the corresponding comoving number density spectrum using \eref{eq:dn_from_beta}.

From the mode functions we construct the Bogolubov coefficient of each species as,
\begin{align}
    \label{eq:betak_sq}
    |\beta_k|^2 
    & = 
    \lim_{t \to \infty} \Bigl[ 
    \frac{\omega_k}{2} | \chi_{\kvec} |^2 
    + \frac{1}{2 \omega_k} | {\dot\chi}_{\kvec} |^2 
    \\ & \qquad \qquad 
    + \frac{\i}{2} \bigl( \chi_{\kvec} {\dot\chi}_{\kvec}^{\ast} - {\dot\chi}_{\kvec} \chi_{\kvec}^\ast \bigr) \Bigr] 
    \;,
    \nonumber
\end{align}
where $\chi_\kvec$ is the canonically-normalized Fourier mode amplitude, and where the dispersion relations are given by
\begin{subequations}
  \label{eq:omegak_sq}
\begin{align}
    {\omega_k^2}^{(\phi)} 
    & = \frac{k^2}{a^2} 
    - \frac{9 H^2}{4} 
    - \frac{3 \dot{H}}{2} 
    + \frac{\tau_0^2}{\bar{\tau}^2} \Vinf''(\bar{\phi}) 
    \\
    {\omega_k^2}^{(\tau)} 
    & = \frac{2 \bar{\tau}^2}{3 \Mpl^2} \partial_\tau^2 V_{\mathrm{KKLT}}(\bar{\tau},0) 
    + \frac{k^2}{a^2} 
    - \frac{9 H^2}{4} 
    \\ & \quad 
    - \frac{3 H \dot{\bar{\tau}}}{\bar{\tau}} 
    - \frac{3 \dot{H}}{2} 
    + \frac{4 \tau_0^2}{\Mpl^2 \bar{\tau}^2} \Vinf(\bar{\phi}) 
    - \frac{\ddot{\bar{\tau}}}{\bar{\tau}} 
    + \frac{\dot{\bar{\tau}}^2}{\bar{\tau}^2} 
    \nonumber \\
    {\omega_k^2}^{(\theta)} 
    & = \frac{2 \bar{\tau}^2}{3 \Mpl^2} \partial_\theta^2 V_{\mathrm{KKLT}}(\bar{\tau},0) 
    + \frac{k^2}{a^2} 
    - \frac{9 H^2}{4} 
    \\ & \quad 
    + \frac{3 H \dot{\bar{\tau}}}{\bar{\tau}} 
    - \frac{3 \dot{H}}{2} 
    + \frac{\ddot{\bar{\tau}}}{\bar{\tau}} 
    - \frac{2 \dot{\bar{\tau}}^2}{\bar{\tau}^2} 
    \per
    \nonumber 
\end{align}
\end{subequations}
The resultant spectrum is then calculated as 
\begin{equation}
  \label{eq:dn_from_beta}
  \dd n = \frac{1}{a(t)^3} \frac{\dd k}{k} n_k 
  \quad \text{where} \quad 
  n_k = \frac{k^3}{2\pi^2} \, |\beta_k|^2
  \com 
\end{equation}
such that $\dd n$ gives the number density of particles at time $t$ with comoving momentum $\kvec$ having $|\kvec|$ between $k$ and $k + \dd k$.
Since $\phi$ and $\tau$ are unstable, $\dd n$ gives their spectra at times after CGPP is completed but before appreciable decays occur. 

\subsection{Spectra and abundances}

\begin{figure}[t]
\centering
\includegraphics[width=0.50\textwidth]{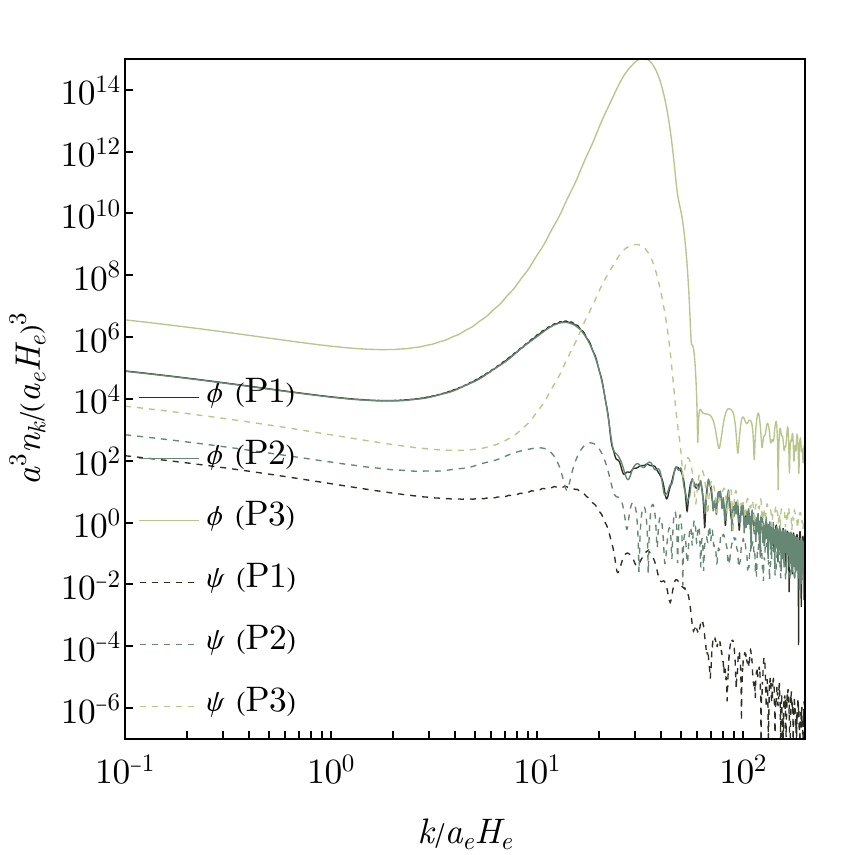}
\caption{\label{fig:spectrum_phi_psi}
    Spectra of gravitationally-produced inflaton $\phi$ and moduli field $\psi$ perturbations.  Both fields are unstable and will decay during reheating. 
}
\end{figure}

\begin{figure}[t]
\centering
\includegraphics[width=0.50\textwidth]{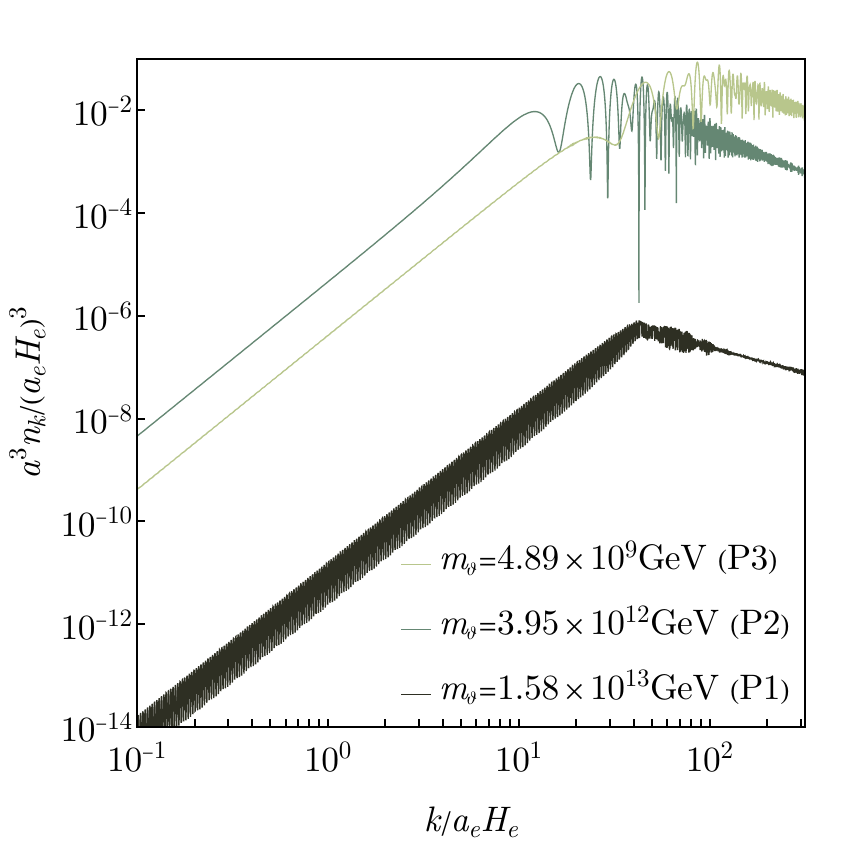}
\caption{\label{fig:spectrum_theta}
    Spectra of gravitationally-produced axion $\vartheta$ perturbations.  If the axion is cosmologically long-lived, it provides a superheavy dark matter candidate.  
}
\end{figure}

Our results appear in \fref{fig:spectrum_phi_psi} and \fref{fig:spectrum_theta}.
In both figures we show several spectra of particles arising from gravitational particle production.
The horizontal axis shows the comoving wavenumber $k$ (or equivalently comoving momentum $p = \hbar k$) in units of the comoving Hubble scale at the end of inflation $a_e H_e$.
Modes with $k / a_e H_e < 1$ leave the horizon during inflation and reenter afterward; modes with $k / a_e H_e > 1$ remain inside the horizon.  
The vertical axis shows the average comoving number density of particles per logarithmic wavenumber (momentum) interval \pref{eq:dn_from_beta} in units of the comoving Hubble volume at the end of inflation. 
A value of $a^3 n_k / a_e^3 H_e^3 = 1$ corresponds to an average density of one particle per Hubble volume at the end of inflation.  

\Fref{fig:spectrum_phi_psi} shows the predicted spectra for the inflaton $\phi$ and the volume modulus $\tau \sim \psi$, corresponding to the models specified by the three parameter sets in \tref{tab:parameters}. 
Towards small $k$, the spectra have a similar shape since the $\psi$ perturbations are sourced by the $\phi$ perturbations.  
Toward large $k$, the spectra exhibit a characteristic ``wiggly'' behavior, which is a consequence of quantum interference effects~\cite{Basso:2022tpd}.
Apart from the oscillations, the average behavior is a power law: $\propto k^{-3/2}$ for $\psi~\mathrm{(P2,P3)}$ and $\propto k^{-9/2}$ otherwise.  
These power laws can be understood to arise from particle production after the end of inflation due to inflaton condensate scattering~\cite{Ema:2018ucl,Chung:2018ayg,Ema:2019yrd}. 
The $k^{-3/2}$ scaling corresponds to $2\to2$ scattering, $\phi \phi \to \psi \psi$, which is only possible for P2 and P3 where $m_\psi < m_\phi$, and the $k^{-9/2}$ scaling corresponds to $2\to3$ scattering $\phi \phi \to 3\phi$ or $3\psi$. 

\Fref{fig:spectrum_theta} shows the predicted spectra for the axion $\theta \sim \vartheta$.
The axion develops a strongly blue-tilted spectrum $\propto k^3$ on large scales.  
This is typical of gravitational production for particles with mass larger than the inflationary Hubble scale; here we have $m_\vartheta / \Hinf = 49.3$, $28.1$, and $112$ for P3, P2, and P1 respectively. 
Quantum fluctuations during inflation, which would contribute to the spectrum at $k/a_eH_e \ll 1$, are suppressed, and gravitational particle production mainly occurs at the end of inflation where causality considerations explain the characteristic $k^3$ scaling. 
The spectra for parameters P1 and P2 oscillate with constant $\ln(k)$ periods for $k/(a_e H_e) \lesssim 10^{1.5}$. 
These large-scale behaviors can be analytically derived, as shown in Appendix A of \rref{Kolb:2023dzp}.
At high-$k$, the spectra exhibit oscillations on top of power law decay, similar to those seen in \Fref{fig:spectrum_phi_psi}.
The power law is $k^{-1}$ for set P1, $k^{-9/2}$ for set P2, and $k^{-3/2}$ for set P3. 
The $k^{-1}$ scaling for set P1 is a consequence of the effective mass term $\bar{\tau}(t)^2 \partial_\theta^2 \VKKLT$, which oscillates at the end of inflation when $\bar{\tau}(t)$ oscillates.  

These spectra are integrated to produce the total number comoving density $a^3 n_\vartheta$ appearing in \tref{tab:cgpp_abundance}.
For parameter sets P2 and P3, the axion mass $m_\vartheta$ is smaller than the inflaton mass $m_\phi$, but is much larger than the inflationary Hubble scale $H_e \sim H_{\mathrm{inf}}$.
In this regime, the abundance can be calculated using an analytical approximation~\cite{Ema:2018ucl,Chung:2018ayg}, which gives $a^3 n_\vartheta / a_e^3 H_e^3 = \frac{3}{8} \frac{1}{16 \pi} \approx 7.5 \times 10^{-3}$. 
For comparison our numerical calculation of the spectrum yields $a^3 n_\vartheta / a_e^3 H_e^3 \approx 3.3 \times 10^{-2}$ for P2 ($a^3 n_\vartheta / a_e^3 H_e^3 \approx 6.4 \times 10^{-2}$ for P3), implying that the analytical approximation works well (it only underestimates the abundance by a factor of $ \lesssim 10$).
The analytical estimate was derived from the resonance associated with an oscillating $\vartheta$ effective mass, which is due to the Hubble terms in \eqref{eq:omegak_sq}. 
For parameter set P1, the axion mass is larger than the inflaton mass, and we find that there is less particle production by several orders of magnitudes. 
This behavior is consistent with the expected exponential suppression of CGPP at high mass; the three parameter sets in \tref{tab:cgpp_abundance} are compatible with a scaling $a^3 n_\vartheta \propto \mathrm{exp}(-4 m_\vartheta / m_\phi)$.

\begin{table}[t]
\caption{Total comoving number density $a^3 n_\vartheta$ of axion $\vartheta$ particles and bounds on the reheating temperature $T_\mathrm{RH}$ for different parameter sets. For $T_\mathrm{RH} \leq T_\mathrm{RH}^\mathrm{(max)}$ the axion relic abundance has $\Omega_\vartheta h^2 \leq \Omega_\mathrm{dm} h^2 \approx 0.12$.  
}
\label{tab:cgpp_abundance}
\begin{center}
  \begin{tabular}{|c||c|c|c|}\hline
    & P1 & P2 & P3  \\ \hline
    $a^3 n_\vartheta~[a_e^3 H_e^3]$ & $8.04\times 10^{-7}$ & $3.28\times 10^{-2}$ & $6.39\times 10^{-2}$ \\ \hline
    $m_\vartheta~[{\rm GeV}]$ & $1.58\times 10^{13}$ & $3.95\times 10^{12}$ & $4.89\times 10^9$ \\ \hline
    $H_e~[{\rm GeV}]$ & $1.32\times 10^{11}$ & $1.32\times 10^{11}$ & $9.67\times 10^7$ \\ \hline
    $m_\phi~[{\rm GeV}]$ & $4.14\times 10^{12}$ & $4.14\times 10^{12}$ & $7.31\times 10^9$ \\ \hline
    $T_{\rm RH}^{(\rm max)}~[{\rm GeV}]$ & $6.27\times 10^9$ & $6.16\times 10^5$ & $3.48\times 10^{11}$ \\ \hline
  \end{tabular} \\
\end{center}
\end{table}

\section{Superheavy Axion Dark Matter}
\label{sec:darkmatter}


The numerical results of the previous section indicate a relatively efficient production of all three particle species: the inflaton $\phi$, the modulus $\tau \sim \psi$, and the axion $\theta \sim \vartheta$. 
In this section we discuss the axion's stability and argue that it may provide a long-lived dark matter candidate.  

\subsection{Axion stability}

At the end of inflation, the inflaton $\phi$ and modulus $\tau$ are expected to decay quickly into Standard Model particles and their superpartners. 
For instance, if these decays proceeded through gravitational-strength interactions then one would expect the inflaton decay rate to be $\Gamma_\phi \sim m_\phi^3 / 8 \pi \Mpl^2$. 
Reheating is completed when the Hubble expansion rate drops down to meet the inflaton decay rate, and the corresponding plasma temperature is $T_\mathrm{RH} \sim m_\phi^{3/2} / \Mpl^{1/2} \sim (5 \times 10^8 \GeV) (m_\phi / 4 \times 10^{12} \GeV)^{3/2}$. 
However, additional non-gravitational interactions typically allow the inflaton to decay much more quickly.
For instance, if reheating occurs almost instantaneously then energy conservation implies a reheating temperature of $T_\mathrm{RH} \sim \sqrt{H_e \Mpl} \sim (2 \times 10^{14} \GeV) (H_e / 1 \times 10^{11} \GeV)^{1/2}$. 
As a result, the energy carried by $\phi$ and $\tau$ is quickly transferred into a plasma of relativistic particles. 

One may ask whether the decays of $\phi$ and $\tau$ can produce axions, which might dominate over the population arising from gravitational particle production. 
For parameter set P1, we have $m_\phi, m_\tau < m_\vartheta$, so decays into axions are kinematically blocked.  
For parameter sets P2 and P3 we have $m_\phi/2 < m_\vartheta < m_\phi$ so two-body decays $\phi \to \vartheta \vartheta$ are still blocked. 
Decays such as $\phi \to \vartheta + \gamma$ are kinematically accessible for sets P2 and P3, and they may occur if there is a photon (or other light particles) in the final state; however, the rate of such decays are more model dependent, since they depend on how the \KKLT{} moduli are coupled to the Standard Model fields. 
Overall, we assume that the decays of $\phi$ and $\tau$ have a negligible impact on the abundance of axions $\theta$. 

In contrast to the rapidly decaying inflaton and volume modulus, the axion 
is stable 
in our simplified model. The conventional decay channel of axions is into Abelian gauge bosons, e.g. photons. However the model as described here contains only non-Abelian gauge fields, which are confined into glueballs at low energies, with a confinement scale $\Lambda \sim( A e^{-\mathfrak{a}\tau_0/\Mpl})^{1/3} \gg m_{\vartheta}$. 
Assuming that $\theta$ does not couple directly to the Standard Model gauge bosons or other dark $U(1)$ gauge bosons, then $\vartheta$ in our simplified model is stable. This can be the case if the SM is realized by a stack of branes wrapping a topologically distinct cycle in a distant region of the compactification. 

Nonetheless one might posit, based on effective field theory arguments, that the axion should be coupled with the SM photon via the conventional interaction,
\begin{equation}
    {\cal L}_{\vartheta \gamma \gamma} = \varepsilon \frac{\alpha_e}{2 \pi f_{\vartheta}} \vartheta F \tilde{F} \;,
\end{equation}
where $\alpha_e \approx 1/137$ is the electromagnetic fine structure constant, $\varepsilon$ is a small parameter to be bounded by the requirement of stability of $\vartheta$, and the mass $m_{\vartheta}$ and decay constant $f_{\vartheta}$ are given by \erefs{eq:mass_formulas}{eq:decayconstant} respectively. This interaction mediates axion decays at a rate given by 
\begin{equation}
    \Gamma_{\vartheta \rightarrow \gamma \gamma} = \varepsilon^2 \frac{\alpha_e^2}{16 \pi^3} \frac{m_{\vartheta}^3}{f_{\vartheta}^2} 
    \per
\end{equation}
In order for the axion to be cosmologically long lived, i.e. to have a lifetime that is much longer than the age of universe today, we impose $\Gamma_{\vartheta \rightarrow \gamma \gamma} \ll H_0 \sim 10^{-33} \eV$, which implies $\varepsilon \ll 10^{-20}$ for parameter set P1. 
Of course if the axion's decay chain includes Standard Model particles, then constraints on the axion's lifetime, arising from an unobserved strong diffuse $\gamma$-ray or cosmic ray background, are much stronger than that from the age of the universe.  
For example, \rref{Das:2023wtk} finds constraints that can be as strong as $\Gamma_{\vartheta\to\gamma\gamma} < 10^{-30} \sec^{-1} \approx 10^{-13} H_0$, depending on the axion's decay products. 
Evidently, it is necessary to forbid (or very strongly suppress) interactions between the axion and the Standard Model gauge fields. Note that QCD axion models also have $\Gamma_{a\to\gamma\gamma} \ll H_0$, thanks in part to the tiny axion mass.

The axion's gravitational interactions also threaten to render it unstable.  
From an effective field theory perspective, one expects the axion $\vartheta$ to have non-renormalizable couplings with gravity.  
These interactions include a gravitational Chern-Simons term 
\ba{
    \mathcal{L}_\mathrm{int} & = - \frac{\vartheta}{\Lambda} R_{\mu\nu\rho\sigma} \tilde{R}^{\mu\nu\rho\sigma} 
    \com
}
where $R_{\mu\nu\rho\sigma}$ is the Riemann tensor, and $\tilde{R}^{\mu\nu\rho\sigma}$ involves an additional contraction with the antisymmetric Levi-Civita tensor. 
One can take the UV cutoff of the EFT to be $\Lambda = \Mpl$.  
This interaction mediates the decay of the axion $\vartheta$ into pairs of gravitons $h$ with a rate given by~\cite{Alonzo-Artiles:2021mym,Ema:2021fdz} 
\ba{
    & \Gamma_{\vartheta \to hh} = \frac{1}{4\pi} \frac{m_\vartheta^7}{\Lambda^2 \Mpl^4} \\
    & \quad \approx 
    \bigl( 1.7 \times 10^{17} \sec \bigr)^{-1} 
    \biggl( \frac{\Lambda}{\Mpl} \biggr)^{\!\!-2} 
    \biggl( \frac{m_\vartheta}{10^{10} \GeV} \biggr)^{\!\!7}
    \per
    \nonumber
}
For the larger axion masses in parameter sets P1 and P2, $m_\vartheta \sim 10^{13} \GeV$, a strong suppression ($\Lambda > 10^{11} \Mpl$) is needed to allow for a cosmologically long-lived axion.   
Alternatively, by fixing $\Lambda = \Mpl$ we find that 
\begin{equation}\label{eq:axion_decay_limit}
    m_{\vartheta} < 8.7 \times 10^{9} \GeV \quad \quad (\Lambda=\Mpl) 
\end{equation}
yields an axion that is cosmologically long lived with respect to decay into gravitons, i.e. $\Gamma_{\vartheta\to hh} < H_0 \approx (4.6 \times 10^{17} \sec)^{-1}$.  
This bound is not satisfied for parameter sets P1 and P2, \eref{eq:axion_decay_limit}, but it is satisfied by parameter set P3, which has $\Gamma_{\vartheta\to hh} \approx 0.02 H_0$.

This model-agnostic bound is slightly stronger the string theory bound reported in \cite{Leedom:2024qgr}, where the decay rate $\Gamma_{\vartheta \rightarrow hh }$ arising from a stack of D7 branes is given by 
\begin{equation}
    \Gamma_{\vartheta\to hh}=\left( \frac{N}{384 \pi^2 f}\right)^2 \frac{m_{\vartheta}^7}{512\pi \Mpl^4}
\end{equation}
where $N = 2\pi/\mathfrak{a}$, and $f=\sqrt{3/2}\Mpl^2/\tau_0$ is defined by the canonical normalization of the axion, Eq.~\eqref{eq:norm}.  For each of the parameter sets studied here, this translates to a stability bound $m_{\vartheta}< 10^{10}$ GeV, in agreement with the estimate \eqref{eq:axion_decay_limit}.

\subsection{Relic abundance}

Assuming they are stable, we calculate the cosmological energy fraction of axion $\vartheta$ particles today as \cite{Kolb:2023ydq} 
\begin{equation}\label{eq:Omegah2}
\begin{split}
	\Omega_\vartheta h^2 
	& = \biggl( \frac{\pi^2 g_{\ast S,0} T_0^3}{270 \Mpl H_{100}^2} \biggr) 
	\biggl( \frac{m_\vartheta H_e T_{\mathrm{RH}}}{\Mpl^3} \biggr)
	\biggl( \frac{a^3 n}{a_e^3 H_e^3} \biggr) \\ 
	& \simeq 
	\bigl( 0.114 \bigr)
	\biggl( \frac{m_\vartheta}{10^{13} \GeV} \biggr)
    \biggl( \frac{a^3 n}{a_e^3 H_e^3} / 10^{-6} \biggr) 
    \\ & \quad 
    \times \biggl( \frac{H_e}{10^{11} \GeV} \biggr)
	\biggl( \frac{T_{\mathrm{RH}}}{10^{10} \GeV} \biggr)
\end{split}
\end{equation}
where $g_{\ast S,0} \simeq 3.91$ and $T_0 \simeq 0.234 \meV$.
This relation assumes that the universe is effectively matter dominated during the epoch of reheating, between the end of inflation and the beginning of radiation domination.  
It also assumes that the axion field is released from Hubble drag at some time before reheating is complete; in other words, $3 m_\vartheta > H_\mathrm{RH}$.  

The measured relic abundance of dark matter is $\Omega_\mathrm{dm} h^2 \simeq 0.12$~\cite{Planck:2018vyg}.  For our numerical results presented in \fref{fig:spectrum_theta} and \tref{tab:cgpp_abundance}, we find that the observed relic density of dark matter is reproduced for $T_\mathrm{RH} = 6 \times 10^9$, $6 \times 10^5$, and $3 \times 10^{11} \GeV$ 
for parameter sets P1, P2, and P3 respectively. 
In other words, there is a wide range of parameters for which superheavy string axions could be produced gravitationally and make up all of the dark matter. 
In fact, the production is so efficient for some cases that a low reheating temperature is needed to suppress the relic abundance by extending the epoch of reheating.

In order to assess whether the required reheating temperature is reasonable, we perform the following estimate. 
If reheating is primarily accomplished through perturbative decays of the inflaton condensate at a rate $\Gamma_\phi$, then the reheating temperature $T_\mathrm{RH}$ may be calculated using the relation $\Gamma_\phi \approx H_\mathrm{RH} \approx \sqrt{ \pi^2 g_\mathrm{RH} T_\mathrm{RH}^4 / 90 \Mpl^2}$.  
Here $g_\mathrm{RH}$ is the effective number of relativistic species in the plasma at temperature $T_\mathrm{RH}$, and we take $g_\mathrm{RH} = 200$ for this estimate. 
If the inflaton decays through gravitational-strength interactions, then its decay rate can be estimated as $\Gamma_\phi \approx \mathcal{N} m_\phi^3 / 16 \pi \Mpl^2$ where $\mathcal{N}$ is a dimensionless factor that depends on the spin and multiplicity of the decay products, and we take it to be $\mathcal{N} = 200$ for this estimate. 
These relations lead to a reheating temperature of 
\begin{align}
    T_\mathrm{RH}
    & \approx \bigl( 5 \times 10^9 \GeV \bigr) 
    \biggl( \frac{m_\phi}{4 \times 10^{12} \GeV} \biggr)^{\!\!3/2}
    \com
\end{align}
where we have fiducialized to the inflaton mass in parameter set P1.  
From \tref{tab:cgpp_abundance} we see that if $T_\mathrm{RH}$ is equal to $T_\mathrm{RH}^\mathrm{(max)} \approx 6 \times 10^9 \GeV$ then axions have the right abundance to make up all of the dark matter ($\Omega_\vartheta h^2 \approx \Omega_\mathrm{dm} h^2$).
These estimates indicate that for parameter set P1 the desired reheating temperature $T_\mathrm{RH} \approx T_\mathrm{RH}^\mathrm{(max)}$ is naturally obtained for perturbative inflaton decay through gravitational-strength interactions.  
For parameter set P2 a lower reheating temperature is needed, which requires interactions to be weaker than gravity, and for parameter set P3 a higher reheating temperature is needed.

Our relic abundance calculation assumes that axions are produced by CGPP and that otherwise their comoving number density remains constant.  
In other words, it's assumed that axions do not come into thermal equilibrium with the primordial plasma, and it's assumed that axion production via inflaton/modulus decay or other channels is negligible in comparison with the population of axions arising from CGPP.  
If the axions have only gravitational-strength interactions with the Standard Model particles, then their rate of production from the thermal bath can be estimated as $\Gamma \sim \sigma(\mathrm{SM} + \mathrm{SM} \to \vartheta \vartheta) \, n_\mathrm{SM} \sim (T^2 / \Mpl^4) (T^3) \sim (T^5 / \Mpl^4)$ corresponding to an $s$-channel graviton exchange. 
This estimate assumes $m_\vartheta < T$, and the rate is even smaller if $m_\vartheta > T$ as a result of Boltzmann suppression.  
Since $\Gamma$ is smaller than the Hubble expansion rate $H \sim T^2 / \Mpl$ for all $T < \Mpl$, we conclude that the axions do not reach thermal equilibrium with the plasma.  
For models with $m_\vartheta > T_\mathrm{RH}$ the Boltzmann suppression ensures that there is negligible axion production by gravity-mediated thermal freeze in \cite{Garny:2015sjg,Garny:2017kha}. 
In general axions may also be produced via scattering and annihilations of the inflaton condensate after inflation \cite{Ema:2018ucl,Mambrini:2021zpp}. 
The production associated with the minimal gravitational interaction between the inflaton and axion fields is already captured by our Bogolubov formalism calculation \cite{Chung:2018ayg,Kaneta:2022gug}, which is manifest in the power-law tail of our spectra toward high $k$. 
If the inflaton-axion interaction were modeled with dimension-6 operators in an effective field theory framework \cite{Lebedev:2022ljz,Lebedev:2022cic}, additional production channels would become available, and one would need to assess their impact on the axion relic abundance on a model-by-model basis.

\section{Discussion and Conclusions}
\label{sec:conclusion}

In this work we have studied string-inspired models of superheavy axions, which provide a theoretically compelling candidate for the cold dark matter since they are naturally very weakly interacting. 
For concreteness we have focused on string axions that arise from the \KKLT{} construction along with a hilltop model of inflation, but we expect that our general conclusions can be extended to other stringy axions and other models of inflation. 
We calculate the spectrum and abundance of axions that arise from cosmological gravitational particle production during inflation and at the end of inflation. 
We perform this calculation using the Bogolubov formalism and by employing numerical integration methods for three benchmark parameter sets, and we compare our numerical results with analytical approximations to the extent that they are available.  
Assuming that the axions are stable, we calculate their cosmological energy fraction today, finding that they could make up all of the dark matter for a reasonable range of reheating temperatures. 

If superheavy string axions are to provide a viable dark matter candidate, then it is essential that they are either stable or at least cosmologically long lived, and our study includes an extensive discussion of this tricky issue. 
Since these particles are superheavy, many final states are kinematically accessible.  
However, since axions do not carry any conserved charges, there are no symmetry arguments to forbid their decay. 
As pseudoscalar fields, axions are odd under spatial inversion, which is a $\mathbb{Z}_2$ parity symmetry. 
However, this alone is not enough to ensure the stability of an axion $\vartheta$ since operators such as $\vartheta R \tilde{R}$ mediate decays $\vartheta \to hh$ into graviton pairs. 
We find that if the axion's mass remains below $\sim 10^{10} \GeV$ then the axion is cosmologically long-lived with respect to decay into graviton pairs via Planck-suppressed operators. 
A systematic assessment of other potential decay channels would require additional model building, which is beyond the scope of this work. 
Needless to say, a string-inspired model in which axions are both superheavy as well as stable would be of great interest to the phenomenology community. 
Even if the superheavy axions are unstable, they may have important cosmological implications for the origin of other cosmological relics, such as the baryon asymmetry of the universe, dark matter, or dark radiation. The phenomenology of decaying superheavy dark matter in string theory has been studied in detail in \cite{Allahverdi:2023nov}.

Through this work we have intended to highlight the heavier side of the string axiverse.  
These superheavy axions are often overlooked, and they have surely received far less attention than their ultra-light cousins. 
Since the phenomenology of superheavy axions has not been thoroughly explored, we hope that this work will encourage future studies. 

Finally, we note that axion-like particles are not the only fields descending from the extra dimensions of string theory. There can also be a spectrum of {\it Kalb-Ramond}-like particles (KRLPs) \cite{Capanelli:2023uwv}. It will be interesting to apply the approach taken here to KRLPS descending from string theory.

\acknowledgments

The authors thank Mustafa Amin, Yuxuan He, Edward Kolb, and Sam S.C. Wong for helpful discussions.  We are grateful to Andrew Frey, Leah Jenks, and Ignacio Quiros Vargas for comments on a draft of this article. This material is based upon work supported (in part: A.J.L.) by the National Science Foundation under Grant No.~PHY-2412797.  E.M. is supported in part by a Discovery Grant from the Natural Sciences and Engineering Research Council of Canada, and by a New Investigator Operating Grant from Research Manitoba. This research was supported by the Munich Institute for Astro-, Particle and BioPhysics (MIAPbP) which is funded by the Deutsche Forschungsgemeinschaft (DFG, German Research Foundation) under Germany´s Excellence Strategy – EXC-2094 – 390783311. \\

\appendix

\section{Supergravity Model of Minimally Coupled KKLT-Inflation}
\label{app:sugra}

In this appendix we provide a simple supergravity model that leads to the Lagrangian in the main text.  
We take $\Mpl = 1$ in this appendix.  
Let the superpotential $W$ and K\"ahler potential $K$ be given by \cite{McDonough:2016der}
\begin{align}
     W & = W_0 + M  X + A\, e^{-\mathfrak{a} T} \\
     K & = 
     -3 \log \left( T+\bar{T} -  \frac{X\bar{X}}{F(\Phi,\bar{\Phi})} \right) 
     - \frac{1}{2} (\Phi - \bar{\Phi})^2
     \com
     \nonumber
\end{align}
where the three chiral superfields are $T = \tau + \ii \theta$, 
$\Phi = (\phi + \ii \varphi)/\sqrt{2}$, and $X = x + \ii \chi$ is nilpotent $X^2=0$ . 
If $F(\Phi,\bar{\Phi}) = f(\tfrac{\Phi + \bar{\Phi}}{\sqrt{2}})$ then the associated scalar potential evaluated at $x = \chi = \varphi = 0$ is given by 
\begin{align}
    V 
    & = \frac{\mathfrak{a}^2 A^2  e^{-2 \mathfrak{a} \tau}}{6 \tau} 
    + \frac{\mathfrak{a} A^2 e^{-2 \mathfrak{a} \tau}}{2 \tau^2} 
    \\ & \quad 
    + \frac{\mathfrak{a} A W_0 e^{-\mathfrak{a} \tau}}{2 \tau^2} \cos(\mathfrak{a}\theta) 
    + \frac{M^2}{12 \tau^2}f(\phi) 
    \com
    \nonumber
\end{align}
where we assume that $W_0$, $M$, $A$, and $\mathfrak{a}$ are real. 
Now choose $f(\phi)=1+(12\tau_0^2/M^2) \Vinf(\phi)$ to arrive at \eref{eq:Lfull}. 
The mass of the sinflaton $\varphi$ can be lifted by allowing $F$ to also depend upon $\Phi-\bar{\Phi}$, and we assume that $m_\varphi$ is much larger than the other masses, so that its dynamics can be neglected during inflation.

\bibliographystyle{JHEP}
\bibliography{refs}

\end{document}